\documentclass[12pt]{article}

\usepackage{amsmath,amssymb}
\usepackage{tabularx}
\usepackage{epsfig}
\usepackage{graphicx}

\begin{document}

\begin{titlepage}

%\rightline{hep-th/yymmnnn}

%\vskip 2cm

\centerline{\large \bf {Bound states between dark matter particles,}}
\centerline{\large \bf {and emission of gravitational radiation}}

\vskip 1cm

\centerline{Grigoris Panotopoulos}

\vskip 1cm

\centerline{ASC, Physics Department LMU,}

\vskip 0.2 cm

\centerline{Theresienstr. 37, 80333 Munich, Germany}

\vskip 0.2 cm

\centerline{email:{\it Grigoris.Panotopoulos@physik.uni-muenchen.de}}

\begin{abstract}
Bound states of two weakly interactive massive particles
are studied. It is assumed that the WIMPonium is formed due to the
gravitational interaction, since the weak interaction can sometimes be 
repulsive. The lifetimes of the spontaneous
emission of gravitational radiation and of the WIMPs annihilation
into a pair of gravitons are computed, and are shown to be many
orders of magnitude larger than the age of the universe.
\end{abstract}

\end{titlepage}

\section{Introduction}

There is accumulated evidence both from astrophysics and cosmology that 
about 1/4 of the energy budget of
the universe consists of so called dark matter, namely a component which 
is non-relativistic and does not
feel the electromagnetic nor the strong interaction. For a review on dark 
matter see e.g.~\cite{Munoz:2003gx}.
Although the list of possible dark matter candidates is long (for a nice 
list see e.g.~\cite{Taoso:2007qk}),
it is fair to say that the most popular dark
matter particle is the LSP in supersymmetric models with R-parity 
conservation~\cite{Feng:2003zu}. The
superpartners that have the right properties for playing the role of 
cold dark matter in the universe are the
axino, the gravitino and the lightest neutralino. By far the most discussed 
case in the literature is the
case of the neutralino (see the classical review~\cite{Jungman:1995df}), 
probably because of the prospects
of possible detection. Up to now the nature of the cold dark matter particle 
is unknown. Several indirect
detection experiments (see e.g.~\cite{experiments}) are looking for 
annihilation products of dark matter in
astrophysical contexts. Recently there has been some interest in annihilation 
of WIMP dark matter via
intermediate long-lived WIMPonium bound states, which can lead to potentially 
very substantial enhancements in
the dark matter annihilation rate~\cite{wimponium}. In these works the 
authors have considered the case in
which the WIMPonium bound states are formed due to the weak interaction. This 
may sound as the only possibility,
since in particle physics gravity is usually ignored, and WIMPs do not 
interact neither strongly nor
electromagnetically. However, it is known from field theory~\cite{fieldtheory} 
that if an interaction is mediated
by a spin one gauge boson it can be either attractive or repulsive, and 
more presicely the force between two
identical particles (e.g.~electron-electron or positron-positron) is 
repulsive, while the force between
non-identical particles (e.g.~electron-positron) is attractive. In the present 
work we wish to study the case in which WIMPonium is formed because of the 
gravitational interaction. It is interesting in 
this case that everything can be computed 
explicitly, and there are certain
expressions for probability rates, cross sections, lifetimes  etc.
As a limited case of applicability, one could 
have in mind the neutralino case (majorana fermions).

A remark is in order here. We shall be considering a framework in which the
weak interaction cannot bind WIMPs together to form WIMPonium. However, the weak
interaction can still mediate transitions and annihilations that most likely
dominate over the gravitational ones computed in the next sections of this 
article. Therefore, in reality we shall be studying a case of academic 
interest in
which the gravitational interactions for the aforementioned processes 
dominate for a WIMP with \emph{no} weak interactions at all, and not a generic
WIMP or neutralino.

Our work is organized as follows. The present article consists of three 
sections. After this introductory section,
we present the analysis and our results in section 2, and finally we conclude 
in the last section.

\section{Analysis}

This is the main section of the present article. In the first subsection we 
recall the quantum mechanical problem for the
$1/r$ potential case, then we compute the transition rate between two 
arbitrary bound states A and B, and finally in the
third subsection we compute the WIMPonium lifetime due to WIMPs annihilation 
into a pair of gravitons.

\subsection{Energy levels and wavefunctions}

In the non-relativistic limit the two WIMPs (particle-antiparticle) system is 
described by the
Hamiltonian
\begin{equation}
H=\frac{p_1^2}{2M}+\frac{p_2^2}{2M}+V(r)
\end{equation}
where $M$ is the WIMP mass, $p_i, 1=1,2$ are the momentum operators, 
and $V(r)$ is the gravitational
potential given by
\begin{equation}
V(r)=-G \frac{M^2}{r} \equiv \frac{\alpha_g}{r}
\end{equation}
with $r$ being the distance between the particle and its antiparticle. 
Ignoring the center-of-mass
motion, which is that of a free particle with mass $2M$, the dynamics of the 
two-body system is
described by the Hamiltonian
\begin{equation}
h=\frac{p^2}{2 \mu}+V(r)
\end{equation}
where $\mu=M/2$ is the reduced mass of the system, and $p$ is the momentum 
operator conjugate to
the
position operator $r$. Because of the spherical symmetry it is convenient to 
use the spherical
coordinate system, in which case one has to solve a one-dimensional problem 
with respect to $r$
\begin{equation}
y''(r)+2 \mu (E-V_{eff}(r))=0
\end{equation}
where the effective potential is given by
\begin{equation}
V_{eff}=V(r)+\frac{l (l+1)}{r^2}
\end{equation}
with $l$ being the angular momentum quantum number. Finally, the wavefunction 
is given by
\begin{equation}
\Psi_{nlm}(r, \theta, \phi)=Y_l^m(\theta, \phi) \frac{y_{nl}(r)}{r}
\end{equation}
with $Y_l^m$ the spherical harmonics, and with the normalization conditions
\begin{eqnarray}
1 & = & \int \: d \Omega |Y_l^m|^2 \\
1 & = & \int_0^\infty \: dr |y|^2
\end{eqnarray}
while the allowed energy levels are given by
\begin{equation}
E_n=\frac{E_1}{n^2}
\end{equation}
where $n$ is the principal quantum number, and $E_1$ is the energy 
corresponding to the ground state.
Later on we shall make use of $E_1, a_0$, and therefore we report here their 
values
\begin{eqnarray}
a_0 & = & \frac{1}{\mu \alpha_g} \\
E_1 & = & -\frac{\mu \alpha_g^2}{2}
\end{eqnarray}
or equivalently
\begin{eqnarray}
a_0 & = & 2 \left ( \frac{m_{pl}}{M} \right)^3 l_{pl} \\
E_1 & = & -\frac{1}{4} \: \left ( \frac{M}{m_{pl}} \right)^4  M
\end{eqnarray}
where $m_{pl} \sim 10^{19}$~GeV is the Planck mass, 
and $l_{pl} \sim 10^{-33}$~cm is the Planck length.

\subsection{Spontaneous emission of gravitational radiation in the quadrupole 
approximation}

In the case of spontaneous emission, a WIMPonium state A makes a transition 
to a state B in the
absence of any incident gravitational wave. In this subsection we shall 
derive the lifetime of the
state A by exploiting the similarities with the electromagnetic theory. Let 
us first recall here
that in atomic transitions the dominant contribution (whenever is allowed by 
the selection rules) in
the spontaneous emission of electromagnetic radiation is the transition in the 
electric dipole
approximation. In the quantum theory of radiation one obtains the following 
expression for the
integrated transition probability between two atomic states A and 
B~\cite{sakurai}
\begin{equation}
\Gamma_{A->B}=\frac{e^2 \omega^3}{3 \pi} \: |\vec{x}_{BA}|^2
\end{equation}
or
\begin{equation}
\Gamma_{A->B}=\frac{\omega^3}{3 \pi} \: |\vec{p}_{BA}|^2
\end{equation}
where $e$ is the electron charge, $\omega$ is the energy of the photon 
emitted, $\vec{x}_{BA}=
\langle B | \vec{x} | A \rangle$
is the matrix element of the position operator between the states A and B, 
and we have defined
$\vec{p} \equiv e \vec{x}$. On the other hand, in the
classical electromagnetic theory the average power for the dipole radiation 
is given by~\cite{griffiths}
\begin{equation}
\bar{P}=\frac{p_0^2 \omega ^4}{12 \pi}
\end{equation}
where $p_0$ is the amplitude of the oscillatory electric dipole
moment of the radiating source, $p(t)= p_0 cos(\omega t)$.
Remarkably enough, one can use the classical result to obtain the
formula for the transition rate $\Gamma_{A->B}$ given above,
avoiding the full quantum mechanical computation. Within the time
interval $\tau=\Gamma^{-1}$, with $\tau$ being the lifetime of the
state A, one photon with energy $\omega$ is emitted, and therefore
we can write
\begin{equation}
\bar{P}=\frac{\omega}{\tau}=\omega \Gamma
\end{equation}
Therefore, the transition rate $\Gamma$ is determined and given by
\begin{equation}
\Gamma=\frac{\bar{P}}{\omega}=\frac{\omega^3}{3 \pi} \: |\frac{p_0}{2}|^2
\end{equation}
which means that from the classical result, the quantum mechanical result 
is obtained provided that
we replace $p_0/2$ by the matrix element $\vec{p}_{BA}$. This is not strange 
at all, if we recall
that in the classical treatment the radiating source is characterized by an 
oscillatory dipole moment
$p(t)=p_0 cos(\omega t)$, while in the quantum theory of radiation the 
interaction Hamiltonian is
written in the form $H_I(t)=H'_I \textrm{exp}(\pm i \omega t)$. Therefore, 
if we write the $cos(\omega t)$
as a sum of two exponentials, $cos(\omega t)=(1/2) (\textrm{exp}(i \omega t)+\textrm{exp}(-i \omega t))$,
we recover the substitutional rule $p_0/2 \rightarrow \vec{p}_{BA}$.

Now let us move to the gravitational radiation case. It is known~\cite{mtw} 
that in the classical
theory of
general relativity the emission of radiation due to electric or magnetic 
dipole vanishes.
The first non-vanishing contribution comes from the quadrupole emission 
mechanism. So, we need the
classical result for the power
for gravitational radiation in the quadrupole approximation, and then we can 
easily obtain the
corresponding quantum mechanical result. In the classical electromagnetic 
theory, the power for
radiation in the quadrupole approximation is given by~\cite{jackson}
\begin{equation}
P_{EM}=\frac{k_c \: \omega^6}{360} \: \sum_{\alpha, \beta} |Q_{\alpha \beta}|^2
\end{equation}
where $k_c=1/(4 \pi \epsilon_0)$ is the constant entering into Coulomb's law 
for the electric force
between two charges, and the quadrupole tensor $Q_{\alpha \beta}$ is defined as
\begin{equation}
Q_{\alpha \beta}=\int \: d^3 \vec{x} \rho(\vec{x}) (3 x_\alpha
x_\beta-|\vec{x}|^2 \: \delta_{\alpha \beta})
\end{equation}
with $\rho(\vec{x})$ being the electric charge density with the property
\begin{equation}
\int \: d^3 \vec{x} \rho(\vec{x}) = Q_{TOT}
\end{equation}
where $Q_{TOT}$ is the total electric charge in space. The classical result 
for gravitational
radiation in the quadrupole radiation looks like the one for electromagnetic 
radiation
\begin{equation}
P_G=\frac{4 G \omega^6}{360} \: \sum_{\alpha, \beta} |I_{\alpha \beta}|^2
\end{equation}
where now $k_c$ is replaced by the Newton's constant $G$, the extra numerical 
factor of 4 is due to
tensor calculus~\cite{mtw}, and $I_{\alpha \beta}$ is the gravitational 
counterpart of $Q_{\alpha \beta}$.
Now we have the mass density with the property
\begin{equation}
\int \: d^3 \vec{x} \rho_M(\vec{x}) = M_{TOT}
\end{equation}
with $M_{TOT}$ being the total mass in space, and the mass quadrupole tensor
\begin{equation}
I_{\alpha \beta}=\int \: d^3 \vec{x} \rho_M(\vec{x}) (3 x_\alpha
x_\beta-|\vec{x}|^2 \: \delta_{\alpha \beta})
\end{equation}
and for the WIMPonium an order-of-magnitude estimation is
\begin{equation}
I_{\alpha \beta} \sim M a_0^2
\end{equation}
Finally, by applying the rule $I_{\alpha \beta}/2 \rightarrow 
I_{BA}^{\alpha \beta}$ we obtain the
transition rate between two WIMPonium states A and B in the quadrupole 
approximation
\begin{equation}
\Gamma_{A->B}^{Gr}=\frac{2 G \omega^5}{45} \: \sum_{\alpha, \beta} 
|I_{BA}^{\alpha \beta}|^2
\end{equation}
where the matrix element $I_{BA}^{\alpha \beta}$ reads
\begin{equation}
I_{BA}^{\alpha \beta}=\langle B | I_{\alpha \beta} | A \rangle
\end{equation}
Taking into account that $G=1/m_{pl}^2$, $\omega \sim |E_1|$, and 
$\langle B | I_{\alpha \beta} | A \rangle
\sim M a_0^2$, we finally obtain the formula
\begin{equation}
\tau_{A->B} \simeq \left( \frac{m_{pl}}{M}  \right )^{15} \: t_{pl}
\end{equation}
where $t_{pl} \sim 10^{-44}$~sec is the Planck time. Assuming a WIMP mass 
$M=10$~TeV we find for the lifetime
\begin{equation}
\tau_{A->B} \simeq 10^{163} t_0
\end{equation}
with $t_0$ being the age of the universe, and we have used the numerical values
\begin{eqnarray}
t_0  & \simeq & 10^{10} yr \\
yr & \simeq & 3 \times 10^7 sec
\end{eqnarray}

\subsection{WIMPonium ground state annihilation into a pair of gravitons}

We assume that the WIMP particles responsible for the cold dark matter in the 
universe are scalar fields.
The available data do not require that WIMPs must be fermions, and as a matter 
of fact there are already
in the literature scalar field candidates for playing the role of dark matter. 
We can mention at least
the case of the axion~\cite{axion}, which has been introduced in order to 
solve the strong CP problem
via the Peccei-Quinn mechanism~\cite{PQ}, and the case of branons that are 
related to the fluctuations of
a D-brane~\cite{branons} on which we are confined according to the brane-world 
idea~\cite{braneworld}.
We choose to work with scalar fields to avoind extra complications from 
spinorial fields, but our final
results should also hold for fermionic WIMPs, since we eventually take the 
non-relativistic limit.

First let us recall how one computes the annihilation rate of the positronium 
ground state into a pair
of photons. Ignoring for one moment that we have a bound state, one needs to 
compute the total
unpolarized cross
section of the process, $e^- e^+ \rightarrow \gamma \gamma$, using standard 
field-theoretic techniques,
then take the non-relativistic limit, and finally take into account that what 
we have is really a bound
state. For that we also need the ground state wavefunction, which has already 
been determined by solving
the Schoedinger's equation. All in all, the inverse of the positronium 
lifetime (ground state) is given
by~\cite{sakurai}
\begin{equation}
\Gamma(n=1, ^1S \rightarrow 2 \gamma)=4 \sigma_{tot}^{unpol} v 
|\Psi_{1s}(\vec{x}=0)|^2
\end{equation}
where the factor of $4$ is due to the spin, namely we have taken into account 
that in the positronium
ground state the single state does the whole job.

Now let us move to the graviton-WIMPs system. According to general relativity, 
if we have a canonical
scalar field coupled to gravity, then the total action describing the model 
contains the
Einstein-Hilbert term for the gravity part plus the appropriate term for the 
scalar field
\begin{equation}
S=\int \: d^4 x \: \sqrt{-g} \: \left (-\frac{R}{2 \kappa^2}+\frac{1}{2}
(\partial \phi)^2-V(\phi)
\right )
\end{equation}
where $R$ is the Ricci scalar, $g$ is the determinant of the metric, 
$\kappa=\sqrt{8 \pi G}$ is the
strength of the gravitational interaction, and $V(\phi)$ is the 
self-interacting potential for the
scalar field. For the discussion to follow we shall consider the case of a 
massive scalar field,
$V(\phi)=(1/2) M^2 \phi^2$. If we write the metric as $g_{\mu \nu}=
\eta_{\mu \nu}+\kappa h_{\mu \nu}$,
it possible to obtain
a usual field theory in Minskowski spacetime, the ``scalar
gravitodynamics'', which describes a massless spin two particle (graviton), 
a massive spin zero particle
(WIMP) interacting with each other. The total Lagrangian takes the 
form~\cite{feynman}
\begin{eqnarray}
S & = & \int \: d^4 x \: \mathcal{L} \\
\mathcal{L} & = & \mathcal{L}[h_{\mu \nu}]+\mathcal{L}[\phi]-\frac{\kappa}{2}
h^{\mu \nu} T_{\mu \nu}+\mathcal{L}[hhh]
+...
\end{eqnarray}
where the first term is for free gravitons, the second term is for free WIMPs, 
the third term
is responsible for the interaction between the gravitons and WIMPs, and the 
rest of the Lagrangian describes
the interaction of gravitons among themselves. Here $T_{\mu \nu}$ is the
energy-momentum tensor for a scalar field
\begin{equation}
T_{\mu \nu}=\partial_{\mu} \phi \partial_{\nu} \phi-\left(\frac{1}{2}
(\partial \phi )^2-V(\phi) \right )g_{\mu \nu}
\end{equation}
Given this interaction term one can derive the Feynman rules for the 
interaction vertices of this
model~\cite{feynman}.
Regarding the free part of the total Lagrangian, one needs to quantize the 
free scalar
field, as well as the free graviton. The scalar field quantization is well 
established in many standard
books on field theory. As for the graviton case, the treatment should be 
similar to quantum
electrodynamics. There are a few differences due to the fact that the photon 
field carries a single
Lorentz index ($A^\mu$, spin one boson, polarization vector 
$\epsilon_{\lambda=1,2}^\mu(k)$), while the
graviton carries two Lorentz indices ($h^{\mu \nu}$, spin two boson, 
polarization vector $\epsilon_{\lambda=1,2}^
{\mu \nu}(k)$). Finally, the Feynman propagator in momentum space for a 
free graviton reads (in the commonly used
De Donder gauge)~\cite{veltman}
\begin{eqnarray}
D_{F}^{\mu \nu \alpha \beta}(k) & = & \frac{P^{\mu \nu \alpha \beta}}
{k^2-i \epsilon} \\
P^{\mu \nu \alpha \beta} & = & \frac{1}{2} (\eta^{\mu \alpha} \eta^{\nu \beta}
+\eta^{\mu \beta} \eta^{\nu \alpha}
-\eta^{\mu \nu} \eta^{\alpha \beta})
\end{eqnarray}
In a similar way to the electromagnetic case and the positronium, we are now 
interested in the process,
$\phi \phi \rightarrow GG$,
with $G$ being the graviton, and we first need to obtain the total unpolarized 
cross section at tree level and in the
non-relativistic limit. Then we can obtain the lifetime of the WIMPonium 
ground state taking into account the corresponding
wavefunction. For the computation of the total cross section for the 
process $1+2 \rightarrow 3+4$ we could start
from the
standard expression for the differential cross section in the center-of-mass 
system given by~\cite{sakurai}
\begin{equation}
\left ( \frac{d \sigma}{d \Omega} \right )_{CM}=\frac{1}{64 \pi^2 s} \: 
\frac{|\vec{p}_f|}{|\vec{p}_i|} \:
|\mathcal{M}|^2 \label{crosssection}
\end{equation}
where $s$ is the Mandelstam's variable, $\vec{p}_i$ and $\vec{p}_f$ are the 
initial and final momenta
respectively, and $\mathcal{M}$ is the Feynman amplitude computed by applying 
the Feynman rules for a given model.
Then we could apply Feynman's rules to compute the invariant amplitude 
$\mathcal{M}$, and finally we could integrate
over the solid angle to obtain the total cross section. However, we do not 
really have to go into this, since we are
only interested in the non-relativistic limit, $M \gg K$, with $K$ being the 
kinetic energy, in which the energy
of the WIMP is $E \simeq M$, and the WIMP momentum is $p \simeq M v$. In this 
limit we can see
from eq. (\ref{crosssection}) that the
cross section
goes like $\sigma \sim 1/v$, and from second order perturbation theory and 
dimensional analysis the cross section
should be
\begin{equation}
\sigma \sim \frac{G^2 M^2}{v}
\end{equation}
where the precise numerical prefactor turns out to be $2
\pi$~\cite{dewitt}. Therefore, in total the WIMPonium lifetime is
given by (since now there is no extra complication related to the
WIMP spin)
\begin{equation}
\tau_{\textrm{WIMPonium}}^{-1}=\Gamma_{\textrm{WIMPonium}}=\sigma_{NR} v 
|\Psi(0)|^2=2 \pi G^2 M^2 \frac{1}{\pi a_0^3}=\frac{1}{4} \:
\left( \frac{M}{m_{pl}} \right )^{10} \: M
\end{equation}
or
\begin{equation}
\tau_{\textrm{WIMPonium}}=4 \: \left( \frac{m_{pl}}{M} \right )^{11} \: t_{pl}
\end{equation}
If the WIMP mass is $M=10$~TeV, then the last formula for the WIMPonium 
lifetime gives
\begin{equation}
\tau_{\textrm{WIMPonium}} \simeq 10^{104} \: t_0
\end{equation}
Therefore, according to our numerical results we conclude that the lifetimes 
(both for WIMP annihilation into a pair
of gravitons and for transitions between bound states in the quadrupole 
approximation) are many orders of magnitude
(in fact astronomically large!) larger than the age of the universe. Since we 
have obtained these results in the non-relativistic
limit, we
expect the same result to be valid for fermionic dark matter particles, 
although our discussion has been focused
on scalar dark matter particles.

\section{Conlusions}

In the present work we have considered bound states formed due to the 
gravitational interaction
between two identical weakly interactive massive particles. These particles 
are the ones responsible
for the cold dark matter in the universe, and
therefore they no not have neither strong neither electromagnetic 
interactions. Furthermore, if the
particle is identical to
its own antiparticle we expect from general field theoretic arguments that 
the weak interaction between
the particle and its
antiparticle should be repulsive. In this case a bound state can be formed 
due to the gravitational
interaction only, which is a
universally attractive interaction. We have assumed that WIMPs
are scalar particles for simplicity, in which case we can easily write down 
the Lagrangian for the
WIMP-graviton system, and determine
the Feynman rules for computing certain processes in quantum gravity. We do 
not worry about the
non-renormalizability of quantum
gravity, since in this work we are only interested in processes at tree level. 
In the non-relativistic
limit the allowed energy
levels and the corresponding wavefunctions of WIMPoniums can be 
computed exactly, and are known from the familiar Hydrogen atom case. We have 
computed
the lifetime of a bound state due to spontaneous emission of gravitational 
radiation in the quadrupole
approximation, as well as
the lifetime of the ground state of WIMPonium due to the WIMPs annihilation 
to a pair of gravitons.
We have determined numerical
values for the lifetimes, which are shown to be many orders of magnitude 
larger than the age of the
universe.

\section*{Acknowledgments}

We wish to thank the anonymous reviewer for his/her comments and suggestions.
This work was supported by project "Particle Cosmology".

\end{document}